# APPLICATION OF THE ASME BOILER AND PRESSURE VESSEL CODE IN THE DESIGN OF SSR CRYOMODULE BEAMLINES FOR PIP-II PROJECT AT FERMILAB[*]

J. Bernardini[†], M. Parise, M. Chen, D. Passarelli
FNAL, Batavia, IL 60510, USA

## Abstract

This contribution reports the design of the main components used to interconnect SRF cavities and superconducting focusing lenses in the SSR Cryomodule beamlines, developed in the framework of the PIP-II project at Fermilab. The focus of the present contribution is on the design and testing of the edge-welded bellows according to ASME Boiler and Pressure Vessel Code. The activities performed to qualify the bellows to be assembled in cleanroom, for operation in high vacuum, cryogenic environments, and their characterization from magnetic standpoint, will also be presented.

## INTRODUCTION

The PIP-II linac [1] will include a total of nine Single Spoke Resonator Cryomodules (SSR CMs), two SSR1 CM [2, 3] and seven SSR2 CM [4]. The SSR section of the linac will accelerate H- ions from 10 MeV to 185 MeV. The PIP-II beam optics design requires that each SSR1 CM contains four focusing lenses (solenoids) and eight SSR1 cavities, and each SSR2 CM contains three focusing lenses and five SSR2 cavities. Each cavity is equipped with one high-power RF coupler, and one tuner, and each solenoid is followed by one Beam Position Monitor (BPM). Cavities, solenoids, and BPMs, together with their interconnecting elements, define the string assembly. The string assembly operates in a high vacuum cryogenic environment and in operation cavities and solenoids need to be aligned within the PIP-II permissible alignment error, reported in Table 1. Therefore, flexible interconnecting elements are needed in the string assembly to compensate for fabrication errors during the mechanical alignment, and to allow for the thermal contraction and expansion of the string's components during cool-down and warm-up cycles.

Edge welded bellows are the element of choice to make such interconnections. They are designed according to ASME Boiler and Pressure Vessel code satisfying all loading scenarios during assembly, cool-down, operation, and warm-up. The bellows are also deigned to be cleanroom compatible, being the SSR strings assembled in an ISO 10 cleanroom [5]. The focus of the present contribution is on the design and testing of the bellows weldments used in the SSR2 string.

[*] Work supported by Fermi Research Alliance, LLC under Contract No. DEAC02-07CH11359 with the United States Department of Energy, Office of Science, Office of High Energy Physics.
[†] jbernard@fnal.gov

Table 1: PIP-II Alignment Requirements

| Cavities | Value |
|---|---|
| Transverse cavity alignment error, mm RMS | <1 |
| Angular cavity alignment error, mrad RMS | ≤10 |
| **Solenoids** | **Value** |
| Transverse cavity alignment error, mm RMS | <0.5 |
| Angular solenoid alignment error, mrad RMS | ≤1 |

## BELLOWS DESIGN

Edge welded bellows have a higher stroke and can be moved laterally more than hydro-formed or expansion-formed bellows of equal length. These features make them a suitable choice for SSR CMs strings [6], where the space is limited and cavities and solenoids are supported from the bottom by the strongback, which stays at room temperature during operations. Lateral movement and stroke are determined by the bellows diameter and number of convolutions. Two different bellows geometries, type 1 - long and type 2 - short, are used among four different weldments in the SSR2 string. Fig. 1 shows the four weldments and Table 2 summarizes the characteristics of short and long bellows.

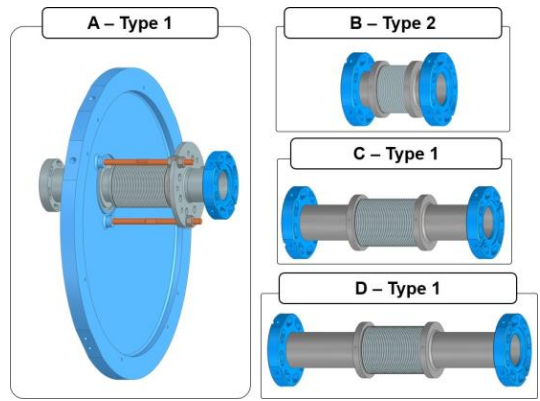

Figure 1: SSR2 edge welded bellows sub-assemblies. A, beamline end flange kit - B, cavity to cavity bellows weldment - C, cavity to BPM bellows weldment - D, cavity to solenoid bellows weldment.

The requirements on the axial stroke and lateral movement are driven by the thermal contraction of the string's components during cool-down / warm-up cycles (up to 1 mm), cavity tuning (up to 2 mm), fabrication errors (up to 4 mm), and installation misalignment. The internal diameter is chosen to be within the beamline aperture, which equals 40 ± 1 mm for SSR1 and SSR2 CMs. The requirement on the life cycle is dictated by the number of cool-down and warm-up cycles that the CM may undergo in a 40 years lifetime (<100).

Table 2: Edge Welded Bellows Characteristics.

| | Free Length [mm] | Min. Length [mm] | Max. Length [mm] | Lateral Disp. [mm] | ID [mm] | OD [mm] | Wall Thickness [mm] | Spring Rate [N/mm] | Life Cycle | Number of convolutions |
|---|---|---|---|---|---|---|---|---|---|---|
| Type 1 | 77 | 30.4 | 84.2 | 5.3 | 39 | 59 | <0.2 | 1.17 | >5'000 | 35 |
| Type 2 | 35 | 20.5 | 42.8 | 2.8 | 39 | 59 | <0.2 | 2.56 | >5'000 | 16 |

The maximum allowable working pressure that the bellows weldment shall sustain is 1.5 bar abs external pressure, and 2.05 bar abs internal pressure. The values of the MAWPs are determined by the pressure setting of the CM's relief valve and the beamline burst disk, respectively. The MAWP that the bellows can sustain is calculated as shown in Eq. 1 (ASME Section VIII, Div. 1 para. UG-101 (m)) [7], where P is the maximum pressure reached in a hydro-static proof test, $S_\mu$ = 463 MPa is the minimum tensile strength at the test temperature specified by ASME Section II, Part D, $E$ = 0.6 is the efficiency of the welded joint (no radio-graphic or ultrasonic inspection is performed on the single-welded butt joint without use of backing strip), and $S_{\mu avg}$ is the average actual tensile strength of the test specimen at the test temperature. The value of $S_{\mu avg}$ is determined by means of tensile tests performed according to ASME Section VIII Div. 1 para. UG-101 (j).

$$MAWP = \frac{B}{4} \times \frac{S_\mu E}{S_{\mu avg}} \qquad (1)$$

According to ASME Section VIII, Div. 1, para. UG-101 (p) vessels having chambers of special shape, subject to collapse, shall withstand a hydro-static pressure test of not less than three times the MAWP without excessive deformation. The total bellows deformation with a 6.15 bar (MAWP × 3) internal pressure gauge is estimated with finite element analysis [8]. Results are shown in Fig. 2 for long bellows.

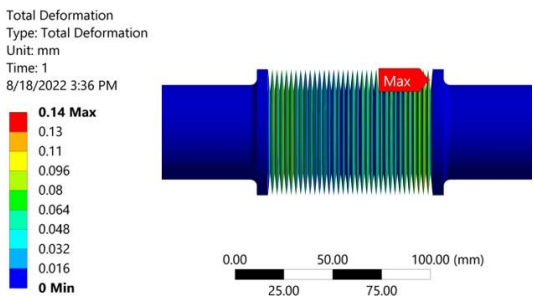

Figure 2: Type A bellows total deformation with 6.15 bar internal pressure gauge

The type 1 bellows used in weldment A in Fig. 1 serves also the purpose to reduce the heat load on the CM 5K and 2K circuits. Fig. 3 shows the temperature distribution along the bellows weldment with (Fig. 3, A) and without (Fig. 3, B) bellows.

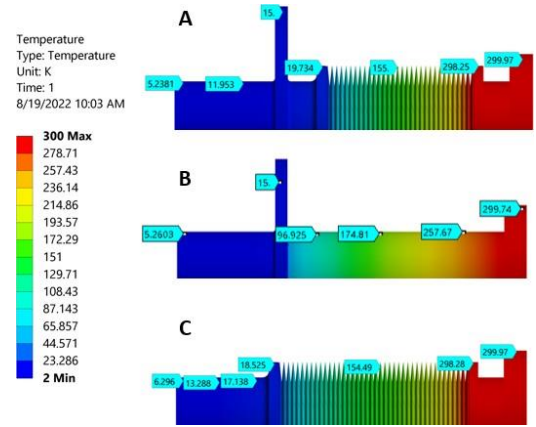

Figure 3: Temperature distribution along the Fig. 2,C bellows weldment. A, type 1 bellows and 15 K thermal intercept - B, no bellows - C, bellows with 45 convolutions and no thermal intercept

The effect of radiation and conduction was considered in the comparative analysis whose results are shown in Fig. 3. The heat reaction on the 15K surface in case A is 91 mW, the heat reaction on the same surface in case B is 455 mW. Fig. 3, C shows the temperature distribution in case the thermal intercept at 15K was removed and a bellows with 45 convolutions was used. The heat reaction in case C is 107 mW, while the heat reaction in case A is 61 mW.

## BELLOWS TESTING

Hydro-static burst test is performed on short and long bellows, pneumatic pressure test and leak checks are performed on the bellows weldments B, C, and D.
The burst test is performed at room temperature, namely the temperature at which the MAWP of 2.05 bar would be reached in case a failure happens during the CM warm-up. The bellows does not burst up to a pressure P = 16.3 bar for long and P = 15.4 bar for short bellows (the hydro-static pressure is hold for 10 minutes). The bellows material is Stainless Steel 316L (UNS S31603 / EN 1.4404), the average actual tensile strength is $S_{\mu avg}$ = 521 MPa. The MAWP resulting from Eq. 1 is 2.27 bar for type A and 2.14 bar for type B bellows. The pneumatic test pressure is 2.4 bar gauge, which is greater than 1.1

× MAWP as required by ASME Section VIII Div. 1, para. UG-100. The test pressure is hold for 10 minutes before performing the leak check. The resulting leak rate is less than $1 \cdot 10^{-9}$ mbar · liter / sec for each bellows weldment.

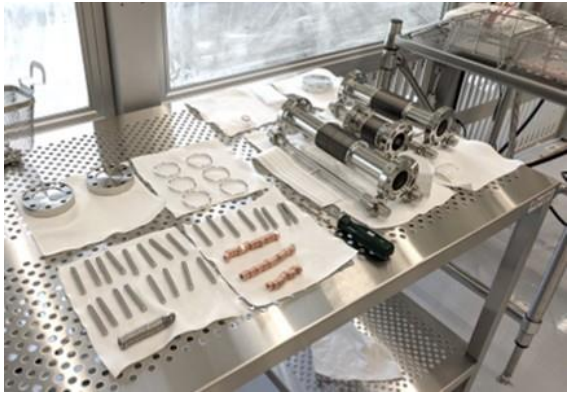

Figure 4: Edge welded bellows weldments and hardware in ISO 10 cleanroom after wet cleaning

The bellows weldments are wet cleaned, cold shocked, and leak checked in a ISO 10 cleanroom. The cleaning procedure includes rinsing in DI water and 99.9% isopropyl alcohol, followed by ultrasonic cleaning, and then dry blowing with filtered nitrogen. A particles counter is used when blowing the outside and inside of the bellows weldment, as shwon in Fig. 5. The counter is set to a 7 second cycle with 75 SLPM pumping speed, and the blowing process takes place till the total number of particulates $> 0.3$ $\mu$m is less than ten in one standard cubic feet of air.

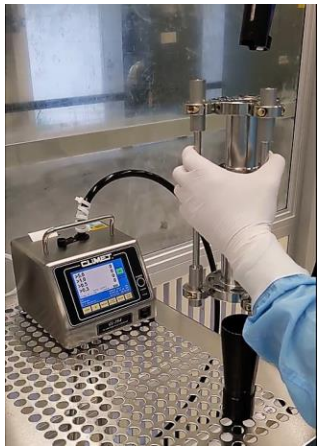

Figure 5: Edge welded bellows weldment dry blowing in ISO 10 cleanroom with filtered nitrogen

Flanges and tubes included in the bellows weldments are electropolished, so that a cosmetic finish and a smoother, easier to clean, surface is obtained. The same cleaning procedure is also performed on the hardware needed for the subsequent tests, as shown in Fig. 4.
A retaining cage and two SS316L flanges are assembled on the bellows weldment by using the same hardware that will be used in the string assembly. Aluminum hexagonal gaskets, SS 316L rods, and Silicone Bronze nuts are used to seal the flanged connections. Nuts are tightened in a star pattern in 3 steps: the tightening torque is 17 Nm in the first step, and 30 Nm in the remaining two steps. The bellows assembly are then cold shocked in liquid nitrogen for three times, and leak checked in between each cold shock. No leak has been detected on all the three bellows weldments with a minimum sensitivity of $2 \cdot 10^{-10}$ mbar · liter / sec.

Table 3: Relative Magnetic Permeability and Residual Magnetic Field. Weldments B,C, and D according to Fig. 1

| Weldment | Location | $\mu_r$ | Residual Magnetic Field [mG] |
|---|---|---|---|
| B | Pipe | 1.000 | <0.1 |
|   | Bellows | 1.007 | 6 |
|   | Flange | 1.039 | 12 |
| C | Pipe | 1.000 | <0.1 |
|   | Bellows | 1.006 | 4 |
|   | Flange | 1.041 | 20 |
| D | Pipe | 1.005 | <0.1 |
|   | Bellows | 1.007 | 5 |
|   | Flange | 1.210 | 42 |

The bellows weldments shall have a relative magnetic permeability $\mu_r < 1.02$ and a residual magnetic field less than 5 mG. Relative magnetic permeability and residual magnetic field after the application of 10 G magnetic field are measured at different locations along the bellows weldments. Results are reported in Table 3. Relative magnetic permeability and residual magnetic field of the flanges are significantly higher than the requirements. The value of $\mu_r$ can be decreased either by using SS 316LN flanges, or by annealing the flanges after machining and welding.

## CONCLUSION

Three samples of SSR2 beamline edge welded bellows weldments were designed, procured, and tested. The bellows weldments were made compliant to ASME Section VIII Div. 1 code by means of hydro-static burst test and pneumatic pressure test. They were cleaned, cold shocked, and leak checked in an ISO 10 cleanroom: all the three weldments successfully passed the leak check and resulted to be easily cleanable. The relative magnetic permeability and residual magnetic field, after the application of 10 G, measured on the flanges of the bellows weldments are higher than allowable. SS316LN flanges shall be used in place of SS316L ones, or annealing shall be performed after machining and welding.